\documentclass[twocolumn,english,reprint, longbibliography, superscriptaddress, breaklinks=true, showkeys, showpacs=false, nofootinbib]{revtex4}
\usepackage[T1]{fontenc}
\usepackage[latin9]{inputenc}
\usepackage{color}
\usepackage{babel}
\usepackage{amsmath}
\usepackage{amssymb}
\usepackage{subfigure}
\usepackage{graphicx}
\usepackage{physics}
\usepackage[unicode=true,pdfusetitle,
 bookmarks=true,bookmarksnumbered=false,bookmarksopen=false,breaklinks=true,pdfborder={0 0 0},backref=false,colorlinks=true]
 {hyperref} 
\makeatletter
\@ifundefined{textcolor}{}
{%
 \definecolor{BLACK}{gray}{0}
 \definecolor{WHITE}{gray}{1}
 \definecolor{RED}{rgb}{1,0,0}
 \definecolor{GREEN}{rgb}{0,1,0}
 \definecolor{BLUE}{rgb}{0,0,1}
 \definecolor{CYAN}{cmyk}{1,0,0,0}
 \definecolor{MAGENTA}{cmyk}{0,1,0,0}
 \definecolor{YELLOW}{cmyk}{0,0,1,0}
}
\pdfoutput=1
\hypersetup{colorlinks=true,citecolor=blue,linkcolor=cyan,urlcolor=blue,filecolor= green, breaklinks=true}
\usepackage{url}
\usepackage{breakurl}
\makeatother

\begin{document}

\title{Quantum simulation of the generalized-entangled quantum eraser \\ and the related complete complementarity relations}

\author{Diego S. S. Chrysosthemos}
\email{starkediego@gmail.com}
\address{Departament of Physics, Center for Natural and Exact Sciences, Federal University of Santa Maria, Roraima Aveneu 1000, Santa Maria, Rio Grande do Sul, 97105-900, Brazil}

\author{Marcos L. W. Basso}
\email{marcoslwbasso@hotmail.com}
\address{Center for Natural and Human Sciences, Federal University of ABC, States Avenue 5001, Santo Andr\'e, S\~ao Paulo, 09210-580, Brazil}

\author{Jonas Maziero}
\email{jonas.maziero@ufsm.br}
\address{Departament of Physics, Center for Natural and Exact Sciences, Federal University of Santa Maria, Roraima Aveneu 1000, Santa Maria, Rio Grande do Sul, 97105-900, Brazil}

\selectlanguage{english}

\begin{abstract}
We utilize IBM's quantum computers to perform a full quantum simulation of the optical quantum eraser (QE) utilizing a Mach-Zehnder interferometer with a variable partially-polarizing beam splitter (VPPBS) at the input. The use of the VPPBS motivates us to introduce the entangled quantum eraser, for which the path information is erased using a Bell-basis measurement. We also investigate the behavior of the wave aspect, i.e., the quantum coherence, as well as the particle character, represented by the predictability and entanglement, as delineated in complete complementarity relations (CCRs). As we show in this article, the utilization of the VPPBS uncover interesting aspects of the QE and CCRs. For instance, we can recover the full wave-behavior by the erasure procedure even when we have only partial knowledge about the path through entanglement.
\end{abstract}


\date{\today}

\maketitle

\section{Introduction}
\label{sec:intro}

The quantum eraser (QE), proposed  by Scully and Dr\"uhl \cite{Scully}, has been carried out  by many authors \cite{Kwiat, Zei, Kim, Walborn, Andersen, Ma} over the last few decades, using several experimental setups, in its normal mode as well as in its delayed choice mode. The main idea of this protocol is that it is possible to manipulate the interplay between the wave-particle duality phenomena using the entanglement between the path of the quantum system (or quanton) in a interferometer and some other degree of freedom. In the context of wave-particle duality phenomena, Bohr's  complementarity principle \cite{Bohr} states that the complete manifestation of the wave property destroys the appearance of the particle property of the quanton, where the wave property is revealed in the visibility of the interference pattern (the swings in the probability graph for the Mach-Zehnder interferometer), while the particle nature is manifested by the which-way information that one can obtain by making a measurement (strong or weak) in one of the arms of the interferometer or by entangling the quanton with another auxiliary quanton and by measuring the auxiliary system. 


\begin{figure}[b]
    \centering
    \includegraphics[scale=0.9]{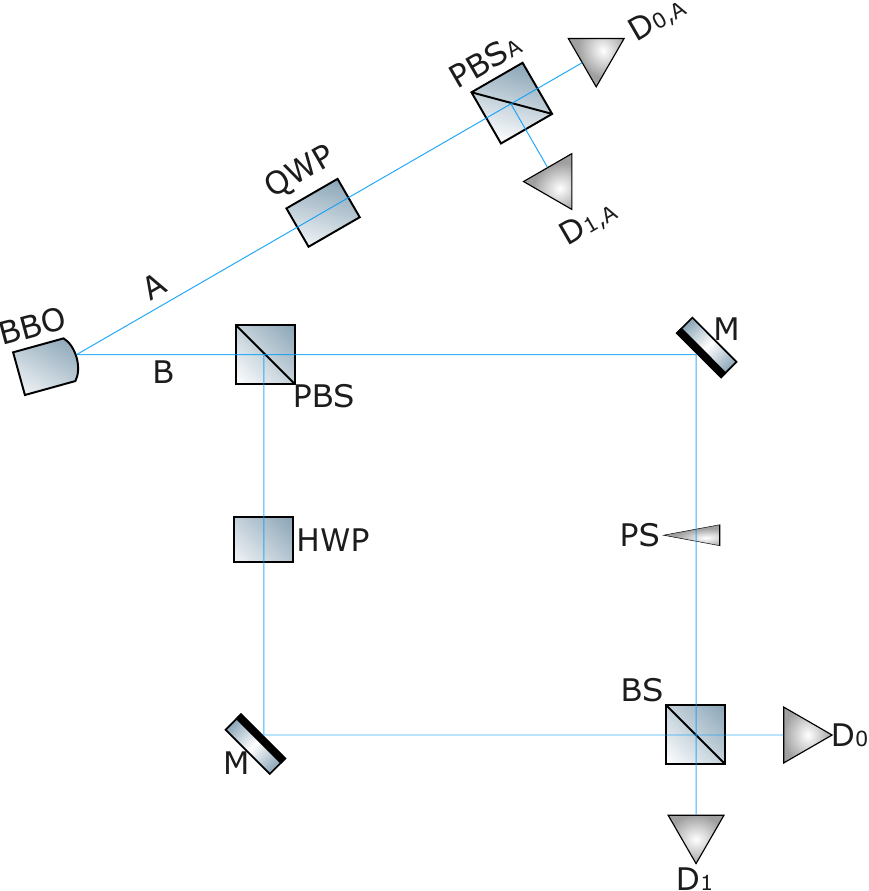}
    \caption{Schematic representation of the quantum eraser. BBO is the non linear crystal of beta barium borate that creates the entangled pair of photons. The quanton $A$ goes to a given path while the quanton $B$ goes through a Mach-Zehnder interferometer. The QWP is the quarter wave-plate, $BS$ is the beam splitter, $M$ stands for the mirrors, HWP is the half wave plate, PS is the phase-shifter and $D_j,D_{jA}$ are the detectors.}
    \label{fig:qeraser}
\end{figure}

Optical quantum eraser experiments \cite{zeilinger} in general involves four qubits. Two photons with two degrees of freedom each (polarization and spatial mode). In this work, the polarization is denoted by $A$ and $B$ and the spatial mode by $A^{\prime}$ and $B^{\prime}$. Regarding notation, $|0\rangle$ represents the horizontal spatial mode as well as the horizontal polarization whereas $|1\rangle$ represents the vertical path as well as the vertical polarization mode. In the QE, which is depicted in Fig. \ref{fig:qeraser}, the entangled pair of photons is created by a non-linear optical crystal, such as beta barium borate (BBO), giving as output the following polarization state: $\left\vert \Psi^{+}\right\rangle_{AB} = 2^{-1/2}  \left(\left\vert 01\right\rangle_{AB} +\left\vert 10\right\rangle_{AB} \right)$. Then, the photon $B$ goes into a Mach-Zehnder interferometer (MZI) and the photon $A$ goes to another region, where it can pass through a quarter-wave plate (QWP) and a polarizing beam splitter (PBS$_A$). The initial state of the system is given by $\left\vert \Psi_{1}\right\rangle  = \left\vert \Psi^{+}\right\rangle _{AB}\left\vert 00\right\rangle_{A^{\prime}B^{\prime}}$. The quanton $B$ passes through a PBS. The PBS let pass the photons with horizontal polarization and reflects photons with vertical polarization. With this, the state of the system right after the PBS is given by $\sqrt{2}\left\vert \Psi_{2}\right\rangle   = \left(  i\left\vert 01\right\rangle _{AB}\left\vert 1\right\rangle _{B^{\prime}}+\left\vert 10\right\rangle _{AB}\left\vert 0\right\rangle _{B^{\prime}}\right)  \left\vert 0\right\rangle _{A^{\prime}},$ where the reflection causes a phase shift of $\frac{\pi}{2} $ and thus a phase shift of $e^{i \frac{\pi}{2}} = i$ in the wave function \cite{Degiorgio, Zeilinger_1981}. At this point, it is worth mentioning that the state $\ket{\Psi_2}$ is genuinely entangled in three degrees of freedom. Later on, this will motivates us to implement what we will call the entangled quantum eraser (EQE). The half-wave plate (HWP) rotates the polarization in the spatial mode wherein it is placed, such that the state after its action is $\sqrt{2}\left\vert \Psi_{3}\right\rangle  = \left(  i\left\vert 01\right\rangle _{AB^{\prime}}+\left\vert 10\right\rangle _{AB^{\prime}}\right)  \left\vert 0\right\rangle_{B}\left\vert 0\right\rangle _{A'}$. So, the HWP disentangles the polarization of the quanton  $B$ from the other degrees of freedom. At this point, one can see that if we measure the horizontal-vertical polarization of $A$, the path information about $B$ inside the MZI is obtained. Now, if we apply the QWP and the PBS$_A$ on quanton $A$, the state of system can be written as $2\sqrt{2}\left\vert \Psi_{4}\right\rangle =-\left(\left\vert 0\right\rangle
_{A}\left\vert 0\right\rangle _{A^{\prime}}\left\vert \psi_{+}\right\rangle
_{B^{\prime}} - \left\vert 1\right\rangle _{A}\left\vert 1\right\rangle
_{A^{\prime}}\left\vert \psi_{-}\right\rangle _{B^{\prime}}\right)\left\vert
0\right\rangle _{B}$, where we defined $\left\vert \psi_{\pm}\right\rangle _{B^{\prime}}=\left(  1\pm e^{i\phi}\right)  \left\vert 0\right\rangle_{B^{\prime}}+i\left(  1\mp e^{i\phi}\right)  \left\vert 1\right\rangle_{B^{\prime}}$. Finally, by making a projective measure in the basis $\{\ket{0}_A, \ket{1}_A\}$ and post-selecting the results, the information about the path of the quanton $B$ is erased and the path coherence of photon $B$ is restored.

Several articles (see e.g. Refs. \cite{Wootters,Yasin,Engle,Bagan,Coles,Hillery,Durr,Englert,Janos,Qureshi,Walborn}) were dedicated to study complementarity relations since the fundamental contributions of de Broglie \cite{deBroglie} and Bohr \cite{Bohr}. Recently, it has been shown that duality inequalities and triality equalities can be derived from the basic properties of the quantum density matrix \cite{Maziero, Marcos, Leopoldo}. This framework has lead to fundamental connections of complete complementarity relations (CCRs) with uncertainty relations \cite{Leopoldo}, entanglement theory \cite{marcos_Emon}, and Lorentz invariance \cite{CCRin}. In Ref. \cite{maleki}, a CCR was applied to quantitatively understand a QE, analog to that of Ref. \cite{Scully}, but also considering partial entanglement of the quanton with the path marker and with an auxiliary system simulating the environment's action.

\begin{figure}[t]
    \centering
    \includegraphics[scale=1]{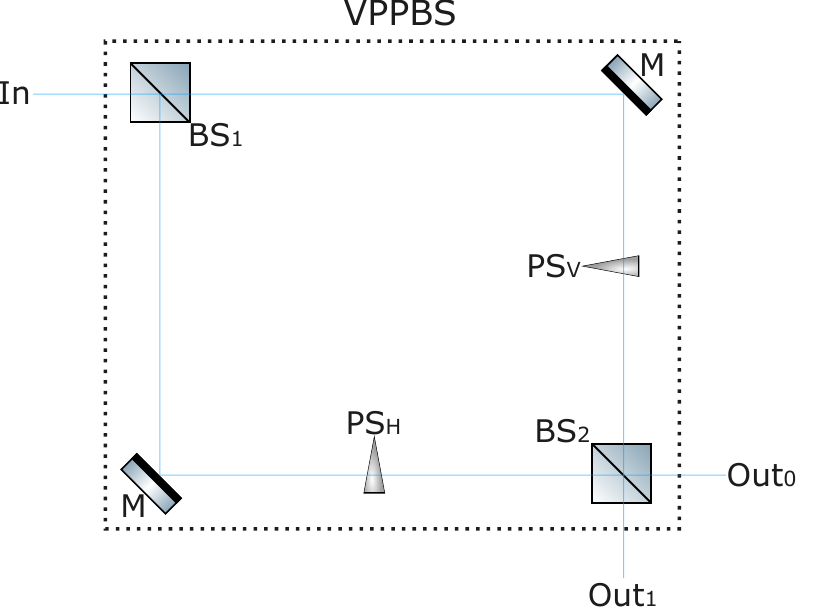}
    \caption{Schematic representation of the variable partially-polarising beam splitter (VPPBS). Everything inside the dotted box represents the VPPBS and there is one input denoted by $In$ where the beam enters and two outputs denoted by $Out_0$ and $Out_1$. BS is the beam splitter, M stands for mirrors, $PS_j$ is the phase-shifter that changes the phase of the polarization $j$, with $j=H,V$.}
    \label{fig:vppbs}
\end{figure}

In this article, we propose a variant to the quantum eraser protocol based on the MZI. We consider a variable partially-polarising beam splitter (VPPBS) that was  recently implemented in Ref. \cite{VPPBS}. With the VPPBS, one can modulate the transmission and reflection of both horizontal and vertical polarization for the quanton $B$, as schematically represented in Fig. \ref{fig:vppbs}. Because of this, as we will see, it is not possible in general to disentangle the polarization of $B$ with the polarization of $A$, as we described above. We propose then an entangled quantum eraser, where instead of measuring just the polarization of $A$, one makes a Bell-basis measurement in the polarizations of $A$ and $B$. Besides, we make a full quantum simulation of the entangled quantum eraser with VPPBS and we investigate complete complementarity relations in this context. It is worthwhile mentioning that, quite recently, the authors in \cite{terno} put forward a experimental setup with three-photon entangled state in which one of the photons is send into a MZI and the output beam splitter of the MZI is controlled by the quantum state of the second photon, which is entangled with a third photon. Therefore, the reduced quantum state of the second photon is undefined, which implements a undefined setting for the MZI. Even though both works uses a tripartite entangled state, the setups are very different.

The remainder of this article is organized in the following manner. In Sec. \ref{sec:VPPBS}, we describe the VPPBS and use it in the quantum eraser experiment, introducing thus the entangled quantum eraser (EQE). Next, in Sec. \ref{sec:ccr}, we investigate complete complementarity relations in the context of the EQE. In Sec. \ref{sec:ibmq}, we use IBM's quantum computers to simulate the EQE and to verify experimentally our theoretical results. Finally, in Sec. \ref{sec:conc}, we give our final remarks.

\section{Quantum eraser with a variable partially polarized beam splitter}
\label{sec:VPPBS}

In this section, we discuss the entangled quantum eraser (EQE) by considering a variable partially-polarizing beam splitter (VPPBS) at the input of the Mach-Zehnder interferometer (MZI). The VPPBS used here is motivated by the experiments carried out recently in Ref. \cite{VPPBS}. Let  $T_j$ and $R_j$ be the transmission and reflection coefficients, respectively, regarding the $j$-polarization, where $j=H,V$ with $H$ and $V$ standing for the horizontal polarization and vertical polarization, respectively. Besides, $T_j$ and $R_j$ are complex numbers such that $|T_j|^2$ is the the probability of a quanton with polarization $j$ being transmitted, while $|R_j|^2$ is the probability of a quanton with polarization $j$ being reflected. Therefore $|T_j|^2 + |R_j|^2 = 1$. The experimental apparatus of the VPPBS is depicted in the Fig. \ref{fig:vppbs}. The coefficients $T_j$ and $R_j$ can be expressed as
\begin{align}
T_{j}(\varphi_{j})  & = e^{i\varphi_{j}/2}\cos\frac{\varphi_{j}}{2}, \label{eq:T}\\
R_{j}(\varphi_{j})   & = ie^{i\varphi_{j}/2}\sin\frac{\varphi_{j}}{2},  \label{eq:R}
\end{align}
with $\varphi_j\in\left[0,2\pi\right]$. It is worthwhile mentioning that the expressions of $T_j$ and $R_j$ are chosen such that $|T_j|^2$ of our manuscript matches the Eq. (7) of Ref. \cite{VPPBS}, where the authors first implemented the variably polarizing beam splitter. However, since we are doing the quantum simulation of VPPBS in the IBMQE, we are free to choose how to implement $T_j$ and $R_j$ in terms of quantum gates given that the relation $|T_j|^2 + |R_j|^2 = 1$ is satisfied and $|T_j|^2$ of our manuscript matches the Eq. (7) of Ref. \cite{VPPBS}. Besides, we must observe that Eqs. (\ref{eq:T}) and (\ref{eq:R}) do not affect our main theoretical findings. 

The unitary transformation performed by VPPBS, $U_{V}^{B^{\prime},B}\left(  \varphi_{H},\varphi_{V}\right)$, on the computational basis for two subsystems, where $B$ is the polarization of the photon $B$ and
$B^{\prime}$ is the path of the photon $B$, is given by
\begin{align*}
U_{V}^{BB^{\prime}}\left(  \varphi_{V},\varphi_{H}\right)  \left\vert
00\right\rangle _{BB^{\prime}} &  =-\left\vert 0\right\rangle _{B}\left(
T_{H}\left\vert 0\right\rangle _{B^{\prime}}+iR_{H}\left\vert 1\right\rangle
_{B^{\prime}}\right) , \\
U_{V}^{BB^{\prime}}\left(  \varphi_{V},\varphi_{H}\right)  \left\vert
01\right\rangle _{BB^{\prime}} &  =-\left\vert 0\right\rangle _{B}\! \left(
-iR_{H} \! \left\vert 0\right\rangle _{B^{\prime}}+T_{H}\! \left\vert 1\right\rangle
_{B^{\prime}}\right) , \\
U_{V}^{BB^{\prime}}\left(  \varphi_{V},\varphi_{H}\right)  \left\vert
10\right\rangle _{BB^{\prime}} &  =-\left\vert 1\right\rangle _{B}\left(
T_{V}\left\vert 0\right\rangle _{B^{\prime}}-iR_{V}\left\vert 1\right\rangle
_{B^{\prime}}\right),  \\
U_{V}^{BB^{\prime}}\left(  \varphi_{V},\varphi_{H}\right)  \left\vert
11\right\rangle _{BB^{\prime}} &  =-\left\vert 1\right\rangle _{B}\left(
iR_{V}\left\vert 0\right\rangle _{B^{\prime}}+T_{V}\left\vert 1\right\rangle
_{B^{\prime}}\right)  .
\end{align*}

\begin{figure}[t]
    \centering
    \includegraphics[scale=0.9]{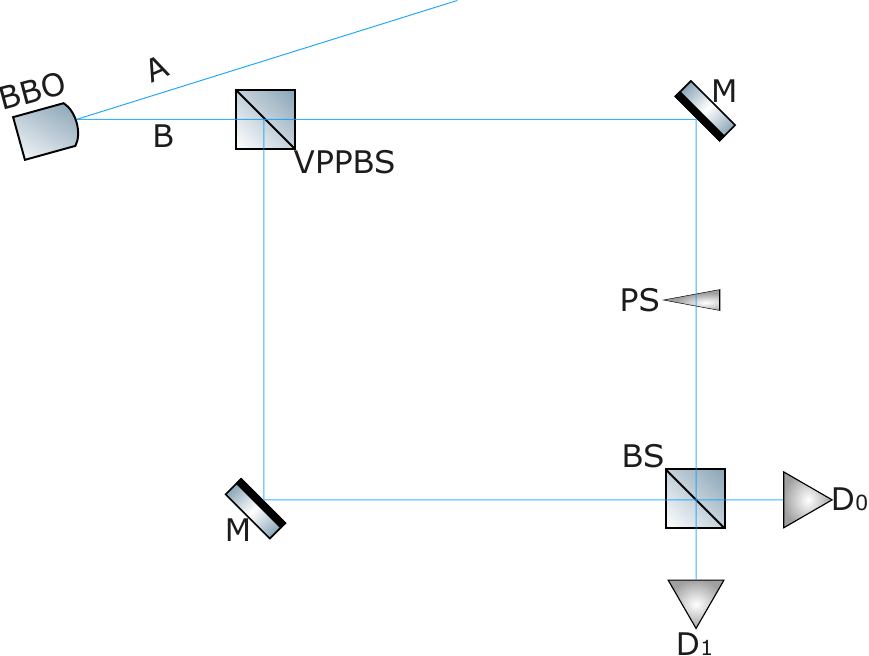}
    \caption{Schematic representation of the  entangled quantum eraser experiment. VPPBS stands for the variable partially-polarizing beam splitter, which is depicted in Fig. \ref{fig:vppbs}. The other optical elements are described in Fig. \ref{fig:qeraser}.}
    \label{fig:qeraser_ent}
\end{figure}

Now, considering the same initial state regarded in Sec. \ref{sec:intro}, $\left\vert \Psi_{1}\right\rangle= \left\vert \Psi^{+}\right\rangle _{AB}\left\vert 00\right\rangle_{A^{\prime}B^{\prime}}$, the state right after the VPPBS in Fig. \ref{fig:qeraser_ent} is given by
\begin{align}
\left\vert \Psi_{2}\right\rangle  = & -\frac{1}%
{\sqrt{2}}\left\vert 01\right\rangle _{AB}\left(  T_{V}\left\vert
0\right\rangle _{B^{\prime}}- iR_{V}\left\vert 1\right\rangle _{B^{\prime}%
}\right)  \left\vert 0\right\rangle _{A^{\prime}} \nonumber\\
&  -\frac{1}{\sqrt{2}}\left\vert 10\right\rangle _{AB}\left(  T_{H}\left\vert
0\right\rangle _{B^{\prime}} + iR_{H}\left\vert 1\right\rangle _{B^{\prime}%
}\right)  \left\vert 0\right\rangle _{A^{\prime}}. \label{eq:psi2}
\end{align}

In the setup for the QE, as seen in Sec. \ref{sec:intro}, the application of the HWP was able to disentangle the polarization state $B$ from the other degrees of freedom. However, with the VPPBS,  given the state in Eq. (\ref{eq:psi2}), this is no longer possible in general since the HWP has the effect of flipping both polarization states, i.e., $\text{HWP} \ket{0} = \ket{1}$ and $\text{HWP} \ket{1} = \ket{0}$. More specifically, after applying the unitary transformation corresponding to the action of the HWP in the state $\ket{\Psi_2}$, the polarizations $A$ and $B$ are generally still entangled. This fact motivates us to consider the entangled quantum eraser (EQE), as described in this section.


For the EQE, let us rewrite Eq. (\ref{eq:psi2}) as
\begin{align}
2\left\vert \Psi_{2}\right\rangle   = & -\left\vert \Psi
_{+}\right\rangle _{AB}\left(  T_{H}+T_{V}\right)  \left\vert 0\right\rangle
_{B^{\prime}}\left\vert 0\right\rangle _{A^{\prime}}\nonumber\\
&  -i\left\vert \Psi_{+}\right\rangle _{AB}\left(  R_{H}%
-R_{V}\right)  \left\vert 1\right\rangle _{B^{\prime}}\left\vert
0\right\rangle _{A^{\prime}} \nonumber \\
&  +\left\vert \Psi_{-}\right\rangle _{AB}\left(  T_{H}%
-T_{V}\right)  \left\vert 0\right\rangle _{B^{\prime}}\left\vert
0\right\rangle _{A^{\prime}}\nonumber\\
&  +i\left\vert \Psi_{-}\right\rangle _{AB}\left(  R_{H}%
+R_{V}\right)  \left\vert 1\right\rangle _{B^{\prime}}\left\vert
0\right\rangle _{A^{\prime}},
\end{align}
where we used the Bell's states $|\Psi_\pm\rangle = 2^{-1/2}\big(|01\rangle\pm|10\rangle\big)$.
For instance, one can see that in the limit where the VPPBS is equivalent to the PBS, i.e., for $T_H=1,\ T_V=0$, we have 
\begin{align}
    2|\Psi_2\rangle = & \big[-|\Psi_+\rangle_{AB}(|0\rangle_{B'}-i|1\rangle_{B'}\big) \nonumber \\
    & + |\Psi_-\rangle_{AB}\big(|0\rangle_{B'}+i|1\rangle_{B'}\big)\big]|0\rangle_{A'}.
    \label{eq:particular}
\end{align}

After the mirrors and phase-shifter in Fig. \ref{fig:qeraser_ent}, the state is turned to
\begin{align}
2\left\vert \Psi_{3}\right\rangle   = & -ie^{i\phi}\left\vert \Psi
_{+}\right\rangle _{AB}\left(  T_{H}+T_{V}\right)  \left\vert 1\right\rangle
_{B^{\prime}}\left\vert 0\right\rangle _{A^{\prime}}\nonumber\\
&  +\left\vert \Psi_{+}\right\rangle _{AB}\left(  R_{H}%
-R_{V}\right)  \left\vert 0\right\rangle _{B^{\prime}}\left\vert
0\right\rangle _{A^{\prime}} \nonumber\\
&  + ie^{i\phi}\left\vert \Psi_{-}\right\rangle _{AB}\left(
T_{H}-T_{V}\right)  \left\vert 1\right\rangle _{B^{\prime}}\left\vert
0\right\rangle _{A^{\prime}}\nonumber\\
&  -\left\vert \Psi_{-}\right\rangle _{AB}\left(  R_{H}%
+R_{V}\right)  \left\vert 0\right\rangle _{B^{\prime}}\left\vert
0\right\rangle _{A^{\prime}}.
\end{align}
Finally, with the action of the last BS, the state of the system is given by
\begin{align}
2\left\vert \Psi_{4}\right\rangle   = &  e^{i\phi}\left\vert \Psi
_{+}\right\rangle _{AB}\left(  T_{H}+T_{V}\right)  \left\vert \ominus
\right\rangle _{B^{\prime}}\left\vert 0\right\rangle _{A^{\prime}}\nonumber\\
&  + \left\vert \Psi_{+}\right\rangle _{AB}\left(  R_{H}%
-R_{V}\right)  \left\vert \oplus\right\rangle _{B^{\prime}}\left\vert
0\right\rangle _{A^{\prime}} \nonumber \\
&  - e^{i\phi}\left\vert \Psi_{-}\right\rangle _{AB}\left(
T_{H}-T_{V}\right)  \left\vert \ominus\right\rangle _{B^{\prime}}\left\vert
0\right\rangle _{A^{\prime}}\nonumber\\
&  - \left\vert \Psi_{-}\right\rangle _{AB}\left(  R_{H}%
+R_{V}\right)  \left\vert \oplus\right\rangle _{B^{\prime}}\left\vert
0\right\rangle _{A^{\prime}}.
\end{align}
As expected, in the limit where the VPPBS is equivalent to the PBS with $T_H=1,\ T_V=0$, one can see that the state above reduces to $\sqrt{2}\left\vert \Psi_{4}\right\rangle =\left(  \left\vert 01\right\rangle _{AB}\left\vert \oplus \right\rangle _{B^{\prime}}+e^{i\phi}\left\vert 10\right\rangle _{AB} \left\vert \ominus\right\rangle _{B^{\prime}}\right)  \left\vert 0\right\rangle _{A^{\prime}}$, where we defined $\vert \oplus \rangle := 2^{-1/2}(\vert 0\rangle+ i \vert 1\rangle)$ and $\vert \ominus \rangle := 2^{-1/2}(\vert 0\rangle- i \vert 1\rangle)$. Then, by doing a BBM on $A$ and $B$ with post-selection, one can recover the wave-behavior of the spatial mode $B'$ in this limit. As for the general case, from the state above, one can calculate the detection probabilities after making a BBM on $A$ and $B$ with post-selection. This calculation will not be displayed here since it's not important for our analysis of the EQE through complete complementarity relations, that will be done in the next section.


\section{A complementarity view on the entangled quantum eraser}
\label{sec:ccr}

Complete complementarity relations (CCRs) arise  in the context of the quantification of Bohr's complementarity principle and allow us to fully characterize a quanton by taking into account not only the predictability and quantum coherence, usually referred as the local aspects of the quanton, but also its quantum correlations with other systems. As discussed in Refs. \cite{Qureshi, Wayhs}, both the predictability and entanglement are linked with the particle behavior of the quanton, such that, when taken together, they can express the path distinguishability in an interferometer, whereas the quantum coherence captures the wave behavior of the quanton. Thus, in the light of CCRs, since we have a tripartite pure entangled quantum system, the path degree of freedom of the quanton $B'$ in  the entangled quantum eraser satisfies the restriction \cite{Marcos}:
\begin{equation}
P_{hs}\left(  \rho_{B^{\prime}}\right)  +C_{hs}\left(  \rho_{B^{\prime}%
}\right)  + S_{ln}\left(  \rho_{B^{\prime}}\right)  =\frac
{d_{B^{\prime}}-1}{d_{B^{\prime}}},\label{eq:ccr}
\end{equation}
where $\rho_{B^{\prime}}$ is the reduced density matrix of the subsystem $B'$ and $d_{B^{\prime}} = 2$ is the dimension of $B'$, $P_{hs}\left(\rho_{B^{\prime}}\right)  = \sum_{j}\left(  \rho_{j,j}^{B'}\right)^{2}-1/d_{B^{\prime}}$ and $C_{hs}\left(  \rho_{B'}\right)  =2\sum_{j\neq
k}\left\vert \rho_{j,k}^{B'}\right\vert ^{2}$ are the predictability and the Hilbert-Schmidt quantum coherence, respectively, 
while  $S_{ln}\left(  \rho_{B^{\prime}}\right) := 1 - \Tr(\rho_{B^{\prime}}^2)$ is an entanglement monotone, as shown in Ref. \cite{marcos_Emon}, that in this case measures the entanglement between $B'$ with the rest of the system as a whole. Besides, as shown in Ref. \cite{CCRin}, Eq.~(\ref{eq:ccr}) is invariant under global unitary operations which implies that such relation remains valid under unitary evolution and, therefore, can be applied in each step of the intereferometer.

\begin{figure*}
\centering
\includegraphics[scale=0.47]{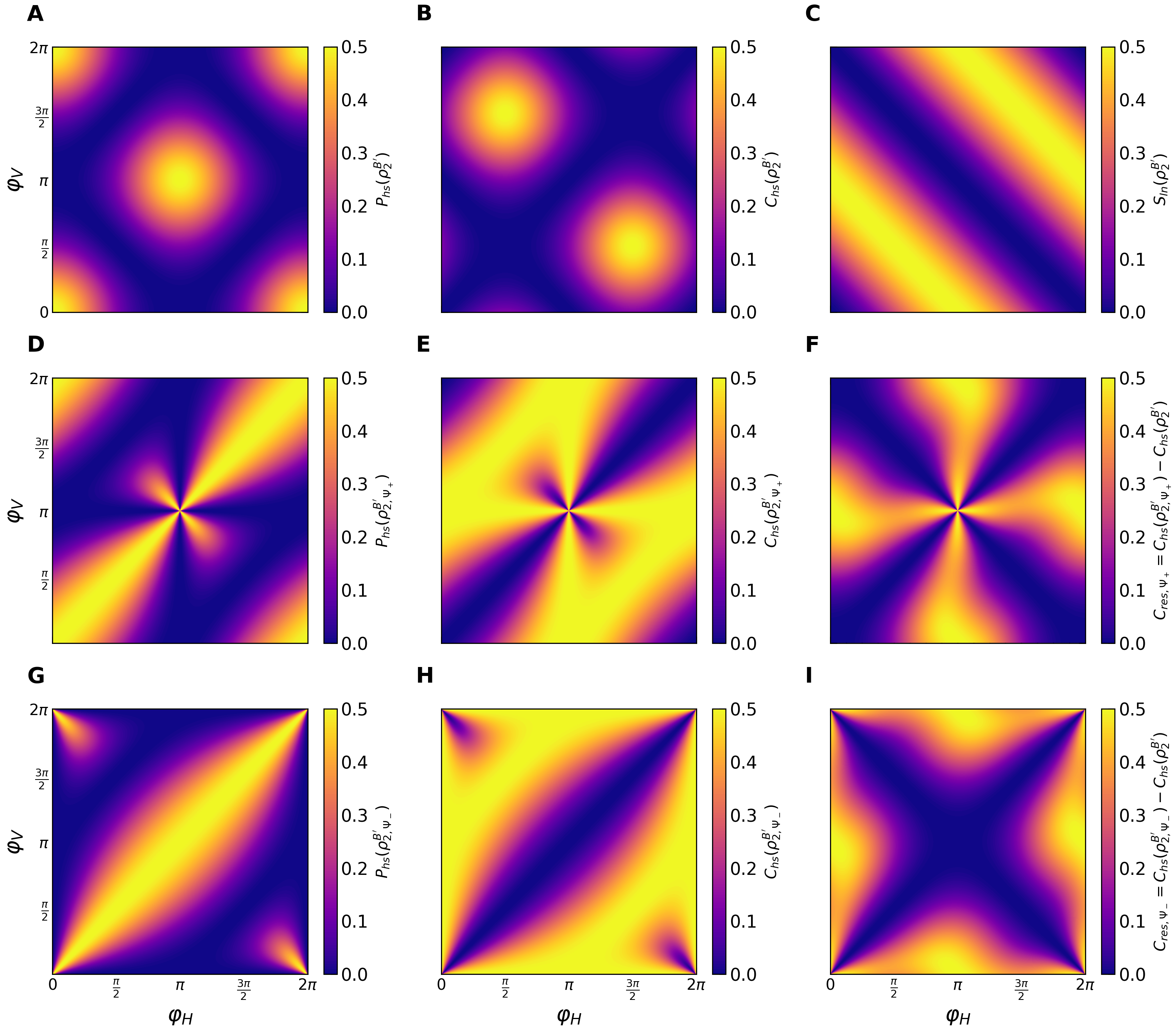}
\caption{Complete complementarity relations before (\textbf{A}-\textbf{C}) and after (\textbf{D}-\textbf{I}) the erasure procedure. Before erasure we regard the state $\rho_2^{B^{\prime}}=\Tr_{ABA'}\big(|\Psi_2\rangle\langle\Psi_2|\big)$ and after the erasure we consider the state $\vert\Psi_{2,\Psi_{\pm}}^{B^{\prime}}\rangle$, where \textbf{D}, \textbf{E} and \textbf{F} correspond to the projection on $\vert \Psi_{+} \rangle_{AB}$ while \textbf{G}, \textbf{H} and \textbf{I} correspond to the projection on $\vert \Psi_{-} \rangle_{AB}$. The functions depend on the angles $\varphi_j$, with $j=H,V$, that are related to the transmission and reflections coefficients (Eqs. \ref{eq:T} and \ref{eq:R}). \textbf{A}. Predictability. The maximum of $P$ occurs  for $T_H = T_V = 0,1$. \textbf{B}. Quantum coherence. The maximum of $C$ occurs when $T_H = 2^{-1}(1+i)$ and $T_V = T_H^{\ast}$, or the complex conjugated of both conditions. \textbf{C}: Entanglement. The maximum of the linear entropy occurs for $T_H = R_V$. \textbf{D}. Predictability. The maximum of $P$ occurs for $|T_H| = |T_V|$. \textbf{E}. Quantum coherence. The maximum path coherence after the erasure, i.e., the maximal path superposition, occurs for $T_H = -R_V$ or when one of the transmission coefficients is zero ($\varphi = \pi$) and the other vary for all the possibles values, except when $T_H = T_V = 0$. In the last case, previous to the erasure, the system can have non null $P$, $C$ and $S$ and all these quantities are converted into maximum path coherence after the erasure, except when $P$ is maximum. When $T_H = -R_V$ previous to the erasure, there is an exchange between $C$ and $S$ and all the entanglement is converted into path coherence after the erasure procedure. \textbf{F}. Restored path coherence. Here is shown the difference between the quantum coherence after and before the Bell-basis measurement, for the post-select state $|\Psi_+\rangle_{AB}$. \textbf{G}. Predictability. The maximum of $P$ occurs for $|T_H| = |T_V|$. \textbf{H}. Quantum coherence. The maximum path coherence after the erasure occurs for $T_H = -R_V$ or when one of the transmission coefficients is one and the other varies for all the possibles values, except when $T_H = T_V = 1$. In the last case, previous to the erasure, the system can have non null $P$, $C$ and $S$ and all these quantities are converted into maximum path coherence after the erasure, except when $P$ is maximum. When $T_H = -R_V$ previous to the erasure, there is an exchange between $C$ and $S$ and all the entanglement is converted into path coherence after the erasure. \textbf{I}. Restored path coherence. Here is shown the difference between the quantum coherence after and before the Bell-basis measurement, for the post-select state $|\Psi_-\rangle_{AB}$.}
\label{fig:PCS_ba}
\end{figure*}

Inside of the MZI, for the global state given by Eq. (\ref{eq:psi2}), we can obtain the quantities involved in the CCR given by Eq. (\ref{eq:ccr}), and we use them to analyse the EQE from the perspective of CCRs. Without measurement and post-selection, the predictability is given by
\begin{align}
P_{hs}\left(  \rho_2^{B^{\prime}}\right)  =&\frac{1}{4}\left(  \left\vert
{T_{H}}\right\vert ^{2}{+}\left\vert {T_{V}}\right\vert ^{2}\right)  ^{2} \nonumber\\
& +\frac{1}{4}\left(  \left\vert {R_{H}}\right\vert ^{2}{+}\left\vert {R_{V}%
}\right\vert ^{2}\right)  ^{2}-\frac{1}{2}. \label{eq:pred}
\end{align}
For the quantum coherence, it follows that
\begin{equation}
C_{hs}\left(  \rho_2^{B^{\prime}}\right)  =\frac{1}{2}\left\vert {T_{H}%
R_{H}^{\ast}-T_{V}R_{V}^{\ast}}\right\vert ^{2}.\label{eq:coh}
\end{equation}
while the  entanglement monotone is given by
\begin{align}
S_{ln}\left(  \rho_2^{B^{\prime}}\right)  =& 1-\frac{1}{4}\left(  \left\vert
{R_{H}}\right\vert ^{2}{+}\left\vert {R_{V}}\right\vert ^{2}\right)
^{2} \nonumber \\
&-\frac{1}{4}\left(  \left\vert {T_{H}}\right\vert ^{2}{+}\left\vert {T_{V}
}\right\vert ^{2}\right)  ^{2} \nonumber \\
& -\frac{1}{2}\left\vert {T_{H}R_{H}^{\ast}-T_{V}R_{V}^{\ast}}\right\vert ^{2}.
\end{align}
These general relations are shown graphically in Fig. \ref{fig:PCS_ba}.(\textbf{A-C}).


Starting from  the state in Eq. (\ref{eq:psi2}), by making a Bell's basis measurement (BBM) with post-selection, the path state $B'$ is reduced to
\begin{align*}
N_{\Psi_{\pm}}\left\vert \psi_{2,\Psi_{\pm}}^{B^{\prime}}\right\rangle 
=&\left(  T_{H}\pm T_{V}\right)  \left\vert 0\right\rangle _{B^{\prime}%
}+i\left(  R_{H}\mp R_{V}\right)  \left\vert 1\right\rangle _{B^{\prime}},
\end{align*}
where the normalization is given by $\left\vert N_{\Psi_{\pm}}\right\vert^{2}=\left\vert T_{H}\pm
T_{V}\right\vert ^{2}+\left\vert R_{H}\mp R_{V}\right\vert ^{2}$.
As we are making a projective measurement in a maximally entangled basis of the systems $A,B$, after the measurements there is no more entanglement between $B'$ and the other degrees of freedom. So, the complete complementarity relation is reduced to
\begin{equation}
P_{hs}\left(  \rho_{2,\Psi_{\pm}}^{B^{\prime}}\right)  +C_{hs}\left(
\rho_{2,\Psi_{\pm}}^{B^{\prime}}\right)  =\frac{1}{2}.\label{eq:ccr2}
\end{equation}
Considering that the density matrix of the corresponding reduced state is given by
\begin{equation}
\rho_{2,\Psi_{\pm}}^{B^{\prime}}=\frac{1}{\left\vert N_{\Psi_{\pm}}\right\vert^{2}}\left[
\begin{array}
[c]{cc}%
\rho_{00}  &  \rho_{01} \\
\rho_{01}^\ast   & 
\rho_{11}
\end{array},
\right]
\end{equation}
where $\rho_{00} = \left\vert T_{H}\pm T_{V}\right\vert ^{2}$, $\rho_{11} = \left\vert R_{H}\mp R_{V}\right\vert ^{2}$ and $\rho
_{01}=-i\left(  T_{H}\pm T_{V}\right)  \left(  R_{H}^{\ast}\mp R_{V}^{\ast
}\right)  $,
the predictability can be written as
\begin{align}
P_{hs}\left(  \rho_{2,\Psi_{\pm}}^{B^{\prime}}\right)  =& \frac
{1}{\left\vert N_{\Psi_{\pm}}\right\vert ^{4}}\left\vert T_{H}\pm
T_{V}\right\vert ^{4}\nonumber\\
&  +\frac{1}{\left\vert N_{\Psi_{\pm}}\right\vert ^{4}}\left\vert R_{H}\mp
R_{V}\right\vert ^{4}-\frac{1}{2}, \label{eq:phspos}
\end{align}
For the quantum coherence, it follows that
\begin{equation}
C_{hs}\left(  \rho_{2,\Psi_{\pm}}^{B^{\prime}}\right)  =\frac{2\left\vert
T_{H}\pm T_{V}\right\vert ^{2}\left\vert R_{H}\mp R_{V}\right\vert ^{2}%
}{\left\vert N_{\Psi_{\pm}}\right\vert ^{4}}. \label{eq:chspos}
\end{equation}
These general relations are shown graphically in Fig. \ref{fig:PCS_ba}.(\textbf{D-I}), where it is also shown the difference between the quantum coherence after and before the erasure procedure (BBM plus post-selection). 

From Fig.~\ref{fig:PCS_ba}, we see that, when $P$ and $S$ are non null and $C = 0$ before the BBM, all the entanglement is converted into predictability by the erasure procedure. On the other hand, when $P$ and $C$ are nonzero and $S = 0$ before the BBM, the erasure procedure does not change these quantities. The more interesting case happens when $C$ and $S$ are the only non null quantities. In this case, characterized by $T_H = -R_V$ and $R_H = - T_V$, there is an initial trade off relation between $C$ and $S$ and all the entanglement is converted into path coherence by the erasure procedure.

The maximum path coherence, for the projection on $\left\vert \Psi_+ \right\rangle$, also occurs after the erasure when one of the transmission coefficients is zero ($\varphi_j = \pi$, for some $j = H,V$) and the other coefficient runs in the interval $[0, \pi) \cup (\pi, 2\pi]$. The point $\varphi_H = \varphi_V = \pi$ should be analysed more carefully. This point corresponds to both transmission coefficients being zero and $\abs{R_j} = 1$ for $j = H, V$. Therefore, after the VPPBS in Fig.~\ref{fig:qeraser_ent}, the state of the system is given by $\ket{\Psi_2} = \frac{i}{\sqrt{2}}(\ket{01}_{AB} -\ket{10}_{AB})\ket{1}_{B'}\ket{0}_{A'}$, representing a state in which the path degree of freedom $B'$ is well defined, with predictability being maximum. Besides, one can see that $B'$ is not entangled with the others degree's of freedom, which implies that the BBM on $A$ and $B$ will not affect the state of $B'$. Hence, one can see that predictability is maximum, while $C = 0$, before and after the BBM, what is expected since the VPPBS reflects the quanton for both polarizations and with probability equal to unity.
The same analysis can be done for the points $\{(\varphi_H, \varphi_V)\} = \{(0,0),(0,2\pi),(2 \pi,0), (2 \pi, 2\pi)\}$.
Another observation to make here is that, when the state of $B'$ after the VPPBS and before the BBM corresponds to a pure separable state, the BBM on $A$ and $B$ will not affect the state of $B'$. This will always happen if the predictability or the quantum coherence of $B'$ after the VPPBS is maximum. For the projection on $\left\vert \Psi_- \right\rangle$, a similar analysis can be done.



Let us proceed with a more careful analysis of some specific cases that appear in this setup. First, the limiting case corresponding to $T_H = 1$ and $T_V = 0$, such that the state $\left\vert \Psi_{2}\right\rangle$ with this set of parameters is given by Eq.~(\ref{eq:particular}).
Thus $P_{hs}\left( \rho_{2}^{B^{\prime}}\right) = C_{hs}\left(  \rho_{2}^{B^{\prime}}\right) = 0$ and $S_{ln}\left(  \rho_{2}^{B^{\prime}}\right) = 1/2$. In other words, we have only path information through entanglement. After a projective measurement on the Bell's basis of $AB$ and post-selection, we obtain one of the following path states
\begin{equation}
\left\vert \psi_{2,\Psi_{\pm}}^{B^{\prime}}\right\rangle =\frac{1}{\sqrt{2}
}\left(  \left\vert 0\right\rangle _{B^{\prime}}\pm i\left\vert 1\right\rangle
_{B^{\prime}}\right).
\end{equation}
Therefore, we recover the wave behavior since $P_{vn}\left(  \rho_{2,\Psi_{\pm}}^{B^{\prime}}\right) = 0$ and $C_{hs} = \left(  \rho_{2,\Psi_{\pm}}^{B^{\prime}}\right) = 1/2$. This is the usual quantum eraser.
As we mentioned above, it is possible to find out a set of parameters for which only two of the following resources $P$, $C$ and $S$ are nonzero. We highlight three situations wherein interesting behaviors are observed after the erasure process. For instance, before the BBM and post-selection, one can see that, when $T_H = T_V^{\ast}$ and $R_H = R_V^{\ast}$,  only $P$ and $C$ are nonzero. In this case, the corresponding state takes the form
\begin{align}
\left\vert \Psi_{2}\right\rangle   = & -\frac{1}{\sqrt{2}}\left\vert
01\right\rangle _{AB}\left(  T_{V}\left\vert 0\right\rangle _{B^{\prime}%
}-iR_{V}\left\vert 1\right\rangle _{B^{\prime}}\right)  \left\vert
0\right\rangle _{A^{\prime}}\nonumber\\
&  -\frac{1}{\sqrt{2}}\left\vert 10\right\rangle _{AB}\left(  T_{V}^{\ast
}\left\vert 0\right\rangle _{B^{\prime}}+iR_{V}^{\ast}\left\vert
1\right\rangle _{B^{\prime}}\right)  \left\vert 0\right\rangle _{A^{\prime}},
\end{align}
and the corresponding path reduced state reads
\begin{equation}
\rho_{2}^{B^{\prime}}=\frac{1}{2}\left[
\begin{array}
[c]{cc}%
2\left\vert T_{V}\right\vert ^{2} & \rho_{01}\\
\rho_{10} & 2\left\vert R_{V}\right\vert ^{2}%
\end{array}
\right],
\end{equation}
where $\rho_{01}=\rho_{10}=i\left(T_{V}R_{V}^{\ast}-T_{V}^{\ast}R_{V}\right)$. So, if $|T_V| = 0$ or $1$ we have maximum predictability. On the other hand, if $|T_V| = |R_V| = 1/\sqrt{2}$ we obtain maximum quantum coherence. After the BBM and post-selection, the path state is collapsed to $\left\vert \psi_{2,\Psi_{\pm}}^{B^{\prime}}\right\rangle =\frac{1}{N_{\Psi_{\pm}}}\left[  \left(  T_{V}^{\ast} \pm T_{V}\right)  \left\vert 0\right\rangle_{B^{\prime}}+i\left(R_{V}^{\ast} \mp R_{V}\right)  \left\vert 1\right\rangle_{B^{\prime}}\right].$
From this reduced state, one can easily see that nothing happens, i.e., the same behavior remains before and after the procedure. For instance, if $|T_V| = 0$ or $1$, $P$ keeps its maximum value and if  $|T_V| = |R_V| = 1/\sqrt{2}$ then $C$ have the same value after and before the erasure procedure.

The second situation is when only $P$ and $S$ are non null. This condition is satisfied with $T_H = T_V = T$. In this case, the state $\left\vert \Psi_{2}\right\rangle$ is reduced to
\begin{align}
\left\vert \Psi_{2}\right\rangle  = & -\frac{1}{\sqrt{2}}T\left(
\left\vert 01\right\rangle _{AB}+\left\vert 10\right\rangle _{AB}\right)
\left\vert 0\right\rangle _{B^{\prime}}\left\vert 0\right\rangle _{A^{\prime}%
} \nonumber\\
& +\frac{i}{\sqrt{2}}R\left(  \left\vert 01\right\rangle _{AB}-\left\vert
10\right\rangle _{AB}\right)  \left\vert 1\right\rangle _{B^{\prime}%
}\left\vert 0\right\rangle _{A^{\prime}},
\end{align}
where the corresponding reduced path density matrix reads
\begin{equation}
\rho_{2}^{B^{\prime}}=\left[
\begin{array}
[c]{cc}%
\left\vert T\right\vert ^{2} & 0\\
0 & \left\vert R\right\vert ^{2}%
\end{array}
\right].
\end{equation}
One can see that if $|T| = 0$ or $1$, there is maximum path information. For $|T_V| = |R_V| = 1/\sqrt{2}$ there is maximum entanglement. After the erasure procedure, the state is given by
\begin{equation}
\left\vert \psi_{2,\Psi_{\pm}}^{B^{\prime}}\right\rangle =\frac{\left(  T\pm
T\right)  \left\vert 0\right\rangle _{B^{\prime}}+i\left(  R\mp R\right)
\left\vert 1\right\rangle _{B^{\prime}}}{\sqrt{\left\vert T\pm T\right\vert
^{2}+\left\vert R\mp R\right\vert ^{2}}},
\end{equation}
and, from Eq.~(\ref{eq:phspos}), one can see that all the entanglement is converted into predictability. This can be interpreted as follows. Once the predictability is related to the \textit{a priori} path-information of the experimentalist \cite{Yasin, Engle}, the information that is still encoded in the entanglement between $B^{'}$ and the rest of the system is then learned by the experimentalist.

The last situation considered here is when only $S$ and $C$ are nonzero before the BBM. This occurs for $T_H = -R_V$ and $R_H = - T_V$ such that the state before the BBM is given by
\begin{align}
\left\vert \Psi_{2}\right\rangle  &  =\frac{1}{\sqrt{2}}\left\vert
01\right\rangle _{AB}\left(  -T_{V}\left\vert 0\right\rangle _{B^{\prime}%
}+iR_{V}\left\vert 1\right\rangle _{B^{\prime}}\right)  \left\vert
0\right\rangle _{A^{\prime}} \nonumber\\
&  +\frac{1}{\sqrt{2}}\left\vert 10\right\rangle _{AB}\left(  iT_{V}\left\vert
1\right\rangle _{B^{\prime}}+R_{V}\left\vert 0\right\rangle _{B^{\prime}%
}\right)  \left\vert 0\right\rangle _{A^{\prime}},
\end{align}
with the corresponding path reduced density matrix given by
\begin{equation}
\rho_{2}^{B^{\prime}}=\frac{1}{2}\left[
\begin{array}
[c]{cc}%
1 & \rho_{01}\\
\rho_{10} & 1
\end{array}
\right],
\end{equation}
where $\rho_{01}=\rho_{10}=i\left(T_{V}R_{V}^{\ast}-R_{V}T_{V}^{\ast}\right)$. Thus, one can see that the maximum entanglement occurs when $|T_V| = 0$ or $1$ and the maximal coherence is obtained for $|T| = |R| = 1/\sqrt{2}$. After the BBM and post-selection the path state takes the form
\begin{align}
\left\vert \psi_{2,\Psi_{\pm}}^{B^{\prime}}\right\rangle   =& \frac{1}{\sqrt
{2}}\left(  T_{V}\mp R_{V}\right)  \left\vert 0\right\rangle _{B^{\prime}%
}\nonumber\\
& -\frac{i}{\sqrt{2}}\left(  T_{V}\pm R_{V}\right)  \left\vert 1\right\rangle
_{B^{\prime}},
\end{align}
such that it is straightforward to see that the corresponding reduced density matrix is given by
\begin{equation}
\rho_{2,\Psi_{\pm}}^{B^{\prime}}=\frac{1}{2}\left[
\begin{array}
[c]{cc}%
1 & \rho_{01}\\
\rho_{01}^{\ast} & 1
\end{array}
\right],
\end{equation}
where $\rho_{01}=i\left(  T_{V}\mp R_{V}\right)  \left(  T_{V}^{\ast}\pm
R_{V}^{\ast}\right)$. 
For all values of $T_V$ and $R_V$, except when coherence is already maximal, it is possible to convert all entanglement into coherence by the erasure procedure. Therefore, we can recover the full wave-behavior by the erasure procedure even when we have only partial knowledge about the path through entanglement. This can be seen analytically by making $T_H = -R_V$ and $R_H = - T_V$ in Eq.~(\ref{eq:chspos}) and using the definition of the coefficients expressed by the Eqs.~(\ref{eq:T}) and (\ref{eq:R}). This is one of the cases that will be experimentally verified in the next section.


\section{Quantum simulation of the entangled quantum eraser on IBMQ}
\label{sec:ibmq}

\begin{figure}[t]
\centering
\includegraphics[scale=0.9]{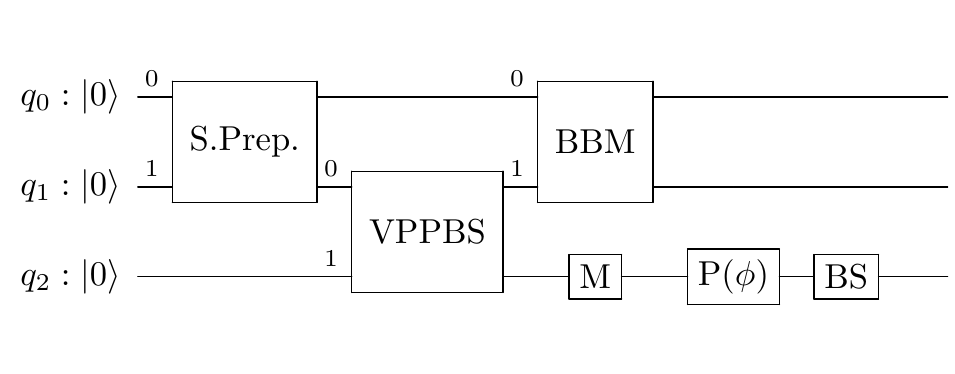}
\caption{The quantum circuit implemented in the IBMQ for simulating the entangled quantum eraser. The qubit $q_0$ is the subsystem $A$, the qubit $q_1$ is the subsystem $B$ and the qubit $q_2$ is the subsystem $B^{\prime}$. The initial states is $\vert 0\rangle$ for all qubits. The $S.\, Prep.$ stands for state preparation of the Bell base state $\left \vert \Psi_+ \right\rangle$. The variable partially-polarizing beam splitter is the VPPBS as the depicted in the Fig. \ref{fig:vppbs}. The $BBM$ performs the measure on the Bell basis on the IBM quantum computers.
For the mirrors, it is necessary to apply the $Z$ gate and then the $Y$ gate. $P(\phi)$ is the phase-shifter and, finally, $BS$ stands for beam splitter.}
\label{fig:qcircuit}
\end{figure}

\begin{figure*}[th]
\centering
\includegraphics[scale=0.55]{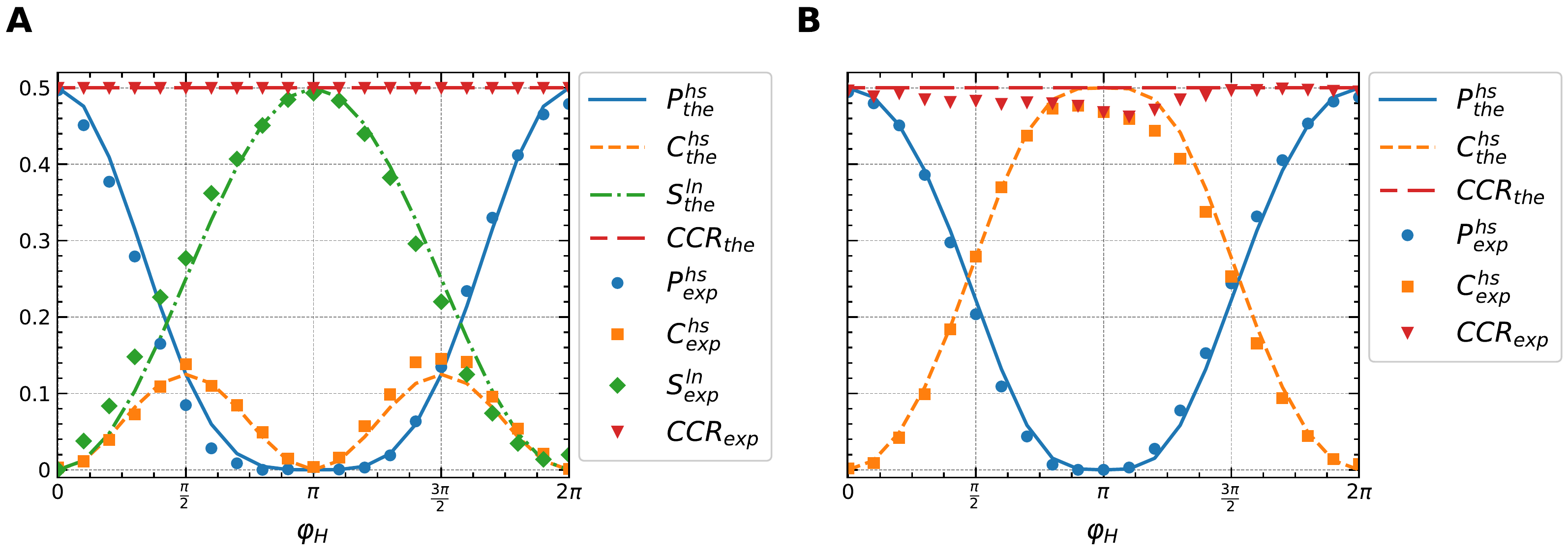}
\caption{Theoretical (lines) and experimental (points) using $\varphi_V = 0$ before (\textbf{A}) and after (\textbf{B}) the erasure procedure on the entangled quantum eraser. \textbf{A}. The functions are computed for the state $\left\vert {\Psi}_{2}\right\rangle$. \textbf{B}. The functions are calculated for the post-selected state $| \psi_{2,\Psi_{+}}^{B^{\prime}}\rangle$. On \textbf{A},  it is possible to see the interchanging between the CCR's functions when changing $\varphi_H$. After the erasure procedure, all the entanglement is converted into predictability or coherence. When $\varphi_H = \pi$ there is maximum path coherence after the erasure. When predictability is maximum, there is no erasure. In the cases where the three functions are non null before erasure, we get partial path coherence restoration.}
\label{fig:vpv=0}
\end{figure*}

\begin{figure*}[th]
\centering
\includegraphics[scale=0.55]{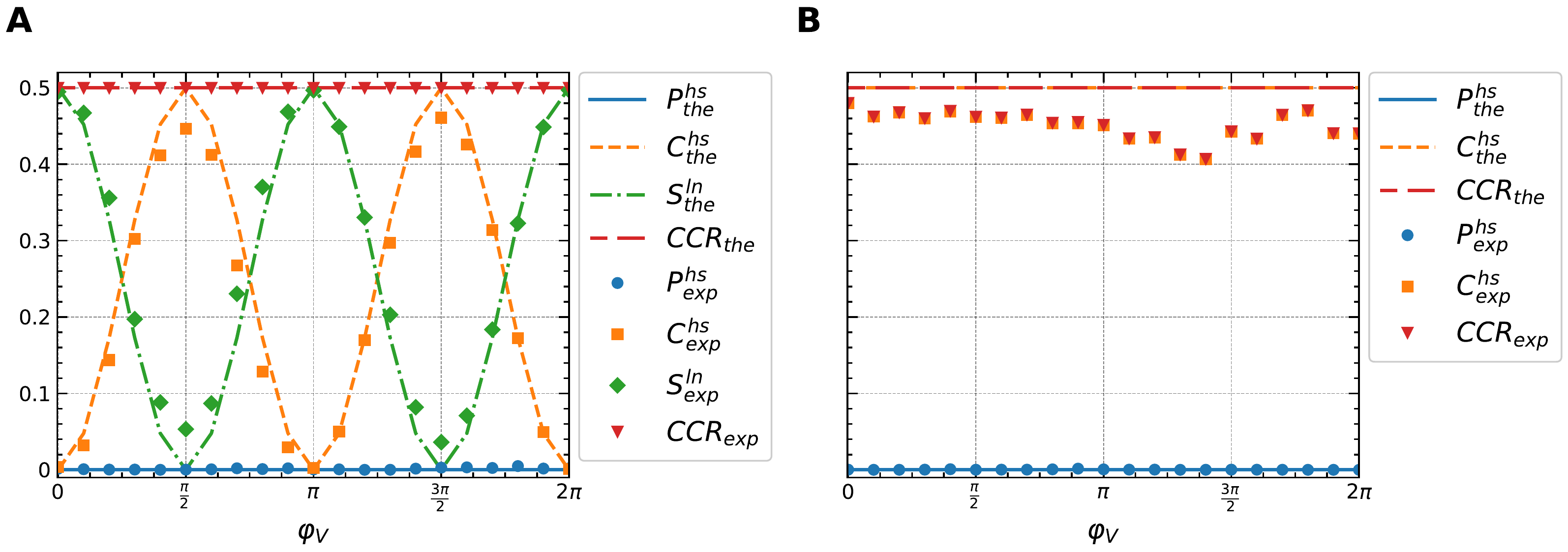}
\caption{Theoretical (lines) and experimental (points) in the case where only $C$ and $S$ are nonzero before the erasure on the entangled quantum eraser setting the relation $\varphi_H = \pi + \varphi_V$ on the VPPBS. \textbf{A}. The functions are regarding to the state $\left\vert {\Psi}_{2}\right\rangle$. 
 \textbf{B}. The functions are regarding the post-selected state $\left\vert \psi_{2,\Psi_{+}}^{B^{\prime}}\right\rangle$. Before the erasure (\textbf{A}) there is an interchanging between $C$ and $S$ and after the erasure (\textbf{B}) all the entanglement is converted into path coherence.}
\label{fig:p=0}
\end{figure*}

In this section, we provide a proof-of-principle experimental verification of our theoretical results using the IBM's quantum computer (IBMQ) \cite{IBMQ}. As seen throughout the article, the EQE is very rich from the perspective of complete complementarity relations. In view of this, we choose some interesting situations for experimental verification. To do this, we implement, as depicted in the Fig. \ref{fig:qcircuit}, the optical elements of the EQE through unitary gates as follows. The qubit $q_0$ is the subsystem $A$, the qubit $q_1$ is the subsystem $B$ and the qubit $q_2$ is the subsystem $B^{\prime}$. The S.Prep. box stands for state preparation, and prepares the following Bell state $\left \vert \Psi_+ \right\rangle$. For S.Prep., we need to apply the Hadamard gate, where $H = \frac{1}{\sqrt{2}}\begin{bmatrix}1&1 \\ 1&-1 \end{bmatrix}$, to the $q_0$ qubit. After that, the controlled X gate (CNOT) ($X$ is the Pauli matrix, the NOT gate), where the control is the qubit $q_0$  and the target is the qubit $q_1$. Finally, the $X$ gate is applied on the qubit $q_1$. The VPPBS, depicted on Fig. \ref{fig:vppbs}, can be constructed as follows. The BS is implemented by $U_{BS}^{B^{\prime}}= SHS$, where $S = \begin{bmatrix}1&0 \\ 0&i \end{bmatrix}$. The mirrors' combined action is implemented using
$
Y^{B^{\prime}}Z^{B^{\prime}} = \begin{bmatrix}
0 & i\\
i & 0
\end{bmatrix},
$
where $Y$ and $Z$ are the usual Pauli matrices. The controlled phase shift for the vertical polarization is implemented through $PS_V = CP_{B,B^{\prime}}\left(  \varphi_{V}\right) =\left\vert 0\right\rangle _{B}\left\langle 0\right\vert \otimes
\mathbb{I}_{B^{\prime}}+\left\vert 1\right\rangle _{B}\left\langle
1\right\vert \otimes P_{B^{\prime}}\left(  \varphi_{V}\right)  $, where the first subscript is the control ($q_2$) and the second is the target ($q_1$). For the horizontal polarization we have the same gate structure, but we need to apply the gate $X$ in the systems $BB^{\prime}$ as follows: $PS_H = \left(  X_{B}\otimes X_{B^{\prime}}\right)  CP_{B,B^{\prime}}\left(  \varphi_{H}\right)
\left(  X_{B}\otimes X_{B^{\prime}}\right)$, with the same structure for the control and the target of the $PS_V$, where $PS = P(\phi) = \ketbra{0} + e^{i \phi} \ketbra{1}$ is the phase gate. The action of the unitary matrix for the VPPBS is given by 
\begin{equation}
U_{V}^{BB^{\prime}}\left(  \varphi_{H},\varphi_{V}\right)  =-\left[
\begin{array}
[c]{cccc}%
T_{H} & -iR_{H} & 0 & 0\\
iR_{H} & T_{H} & 0 & 0\\
0 & 0 & T_{V} & iR_{V}\\
0 & 0 & -iR_{V} & T_{V},
\end{array}
\right]  
\end{equation}
where $T_j$ and $R_j$ are given by the Eqs. (\ref{eq:T}) e (\ref{eq:R}). Finally, the BBM gate performs the measurement on the Bell's basis. In order to implement the BBM gate, we need to apply the CNOT with the control as $q_0$ and the target as $q_1$ and then apply $H$ on the qubit $q_0$. The BBM action allows us to perform the Bell basis measurement on IBM quantum computers.

There are many possible experiments that could be done depending on different sets of parameters used in the VPPBS. We performed experiments in only two situations and the results are presented in Figs. \ref{fig:vpv=0} and \ref{fig:p=0}. We believe that these experiments summarize some of main aspects of the EQE introduced in this article. For these experimental results, we used IBM's Quantum Experience Lima quantum chip.

Although we present the complete experimental apparatus of the EQE (Fig. \ref{fig:qcircuit}), for the production of experimental results we performed two steps for each one of them. In the first step, we performed the state tomography only for the qubit $q_2$ before the mirror. In this way we obtain the path density operator for the photon $B$ and with this we can calculate the functions of Eq. (\ref{eq:ccr}) before the erasure. In the second step, we insert the BBM and perform the state tomography on the qubits $q_0$ and $q_1$ after the BBM and on the qubit $q_2$ before M. Thus, we obtain the density operator, but now for the whole system, and we project it in the desired state, in this case for Bell base state $\vert \Psi_+\rangle$. We normalize the obtained path matrix and so we can calculate the functions of Eq. (\ref{eq:ccr2}) after the erasure procedure. Furthermore, we used the Qiskit tools for measurement error mitigation \cite{qiskit}, which improved substantially the experimental results.

For the first experiment considered, in Fig. \ref{fig:vpv=0}, we fix the $\varphi_V = 0$ and vary $\varphi_H$, regarding to the state $|\Psi_2\rangle$. Before the erasure (\ref{fig:vpv=0}.\textbf{A}) there is a interchange between all the CCR's functions for different values of $\varphi_H$. After the erasure procedure (Fig. \ref{fig:vpv=0}.\textbf{B}) all the entanglement is converted into predictability or coherence with respect to the post-selected state $\left\vert \psi_{2,\Psi_{+}}^{B^{\prime}}\right\rangle$. When $\varphi_H = \pi$, there is maximum path coherence after the erasure. When the predictability is maximum there is no erasure,  while in the cases where the three functions are non null initially, we obtain partial erasure. 

In the next case, Fig. \ref{fig:p=0}, we choose the case discussed at the end of Sec. \ref{sec:ccr}. Here there is a relationship between the two angles given by $\varphi_H = \pi + \varphi_V$. As can be seen, the predictability is always null and we only have the interplay  between C and S before the erasure procedure. After the erasure all entanglement is converted to path coherence.

\section{Conclusions}
\label{sec:conc}

In this article, we have investigated the quantum eraser for a Mach-Zehnder interferometer (MZI) with a variable partially-polarizing beam splitter (VPPBS) at the input. We showed that because of the VPPBS, one cannot, in general, untangle the polarization of the quanton going through the MZI from its path and the polarization of the other photon. This motivated us to introduce the entangled quantum eraser (EQE), for which the path information erasure is performed via a Bell's basis measurement followed by post-selection. We studied this system from the complete complementarity relations perspective, elucidating the machinery of the quantum eraser as a function of the parameters of the VPPBS. We showed that although the EQE typically increases the path coherence, it is not in general maximal at the end of the protocol. As we found, this is so because the initial entanglement can also be transformed into predictability of the post-measurement state. For some specific parameters of the VPPBS, we have discussed in details situations where only two of the three CCR functions are non null. When only $P$ and $C$ are nonzero before the erasure, nothing happens after the erasure procedure. When only $P$ and $S$ are non-null before the erasure, after the erasure we are left only with path information through $P$. Finally, when we have only $C$ and $E$ different from zero initially (this last case was verified experimentally as depicted in Fig. \ref{fig:p=0}), all the entanglement is converted into path coherence after the erasure. 
Finally, we used IBM's quantum computers as a quantum simulator of the optical EQE. We applied it for some interesting cases that illustrate our new setup. Our simulation results agreed quite well with the theoretical predictions. Even with the high noise rates of nowadays quantum computers, our experimental results matched fairly well with the theory. We leave as an open problem the experimental implementation of EQE in the optical setup, once in this scenario one would have to perform a Bell-basis measurement in the polarization degree of freedom of both photons without interfering with the path state of one of them nor affecting the system global state.


\begin{acknowledgments}
This work was supported by the Coordination for the Improvement of Higher Education Personnel (CAPES), process 88887.649600/2021-00, by the 
S\~ao Paulo Research Foundation (FAPESP), Grant No.~2022/09496-8.3, by the National Council for Scientific and Technological Development (CNPq), process 309862/2021-3, and by the National Institute for the Science and Technology of Quantum Information (INCT-IQ), process 465469/2014-0.
\end{acknowledgments}


\end{document}